\documentclass{elsart}
\newcommand{\sst}{\scriptscriptstyle}
\usepackage{amssymb,graphicx,bbm}
\begin{document}
\begin{frontmatter}
\title{Teleportation of bipartite states using a single entangled pair}
\author{Mary M. Cola and Matteo G. A. Paris}
\address{Dipartimento di Fisica and INFM, Universit\`a di
Milano, Italia.}
\date{\today}
\begin{abstract}
A class of quantum protocols to teleport bipartite (entangled)
states of two qubits is suggested. Our schemes require a single
entangled pair shared by the two parties and the transmission of
three bits of classical information, as well as a two-qubit gate 
with an additional qubit at the receiver' location. Noisy quantum 
channels are considered and the effects on both the teleportation 
fidelity and the entanglement of the replica are evaluated.
\end{abstract}
\maketitle
\end{frontmatter}
\section{Introduction}\label{s:intro}
Quantum mechanics allows the phenomenon of quantum teleportation.
An unknown quantum state is destroyed at a sending place (Alice)
while a perfect replica appears at a remote place (Bob) via the
combined action of a quantum and a classical channels
\cite{bennett93}. Teleportation has been demonstrated for the
polarization state of a photon \cite{Zeilinger98}, the state of a
trapped ion \cite{recent} and, in the continuous variable regime,
for the set of coherent states of a single-mode radiation field
\cite{Furusawa}. Perfect quantum teleportation requires a
maximally entangled state as quantum channel. Nevertheless, the
protocol also works for nonmaximally or even mixed quantum channel
\cite{popescu94}, with fidelity of quantum teleportation still
being better than any classical communication procedure.
\par
Initially, attention was mostly focused to teleportation of
single-system quantum states, either of finite-dimensional N-level
systems \cite{bennett93,Stenholm98}, or of single-mode continuous
variables systems \cite{Vaidman94,Ralph98}. More recently,
attention has been devoted to teleportation of states of
multipartite (possibly entangled) systems. Notice that the
sufficient and necessary conditions to induce entanglement on two
remote qubits, by means of their respective linear interactions
with a two-mode driving field, has been recently investigated
\cite{kim1}. Direct transmission of an entangled state was
considered in a noisy environment \cite{schumacher96}, and the
possibility to copy pure entangled states was studied
\cite{Koashi98}. A straightforward generalization of single-body
teleportation \cite{LeeKim,Ikram}, shows that any state of a
$N$-partite system can be teleported using $N$ maximally-entangled
pairs as quantum channel and the transmission of classical
information. In addition conclusive teleportation in
$d$-dimensional Hilbert space for partially entangled quantum
channel has been formulated \cite{kim2}. Moreover a scheme to
teleport two-particle entangled states has been proposed, using
three-particle entangled states, either of GHZ- \cite{Shi} or
W-type \cite{Shi2,Cao}, as quantum channel. These schemes can be
also generalized to teleport an unknown three-particle entangled
state from a sender to any one of $N$ receivers.
\par
In this paper, we focus our attention to teleportation schemes for
bipartite (entangled) states of two qubits. Our goal is to weaken
the requirements for the quantum channel, {\em i.e.} to devise
teleportation protocols that work with a reduced amount of
entanglement shared between the two parties. As we will see, there
is a whole class of protocols realizing this task. In our schemes,
an unknown bipartite state can be perfectly teleported, sharing a
single EPR pair, once one has at disposal the following
ingredients: i) a two-qubit Bell measurements; ii) an Hadamard
transformation and a single-qubit measurement at the sender's
side; iii) an additional qubit and a set of two-qubit unitary
transformations at the receiver's location.
\par
The paper is structured as follows. In Section \ref{s:cnot} the simplest
example of our class of teleportation protocols is described in details,
whereas the general scheme is addressed in Section \ref{s:uf}.
In Section \ref{s:noise} we analyze how our protocols work with
noisy quantum channels, and evaluate teleportation fidelity as well as
entanglement of the teleported state. Section \ref{s:outro} closes the
paper with some concluding remarks.
\section{Teleportation of bipartite states using a single entangled
pair} \label{s:cnot} 
\begin{figure}[h]
\begin{center}
\includegraphics[width=0.7\textwidth]{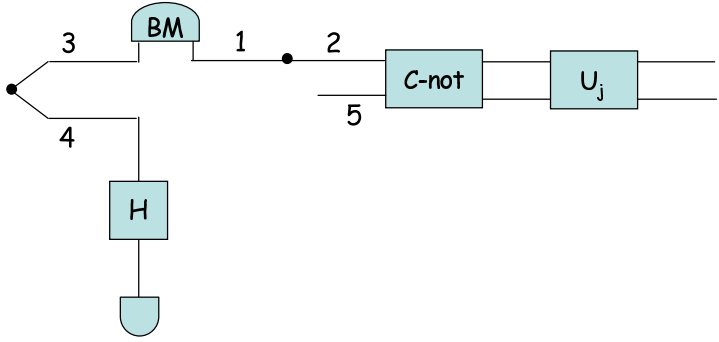}
\end{center}
\caption{Schematic diagram of the suggested teleportation
protocol. Alice and Bob share a single EPR pair (qubits 1 and 2).
Alice performs a Bell measurement (BM) on qubits 1 and 3 and,
after an Hadamard transformation (H), measure qubit 4. Then she
sends to Bob the results of both measurements (overall three bits
of classical information). Bob introduces the additional qubit 5,
and performs a C-not transformation on qubits 2 and 5. Finally, he
performs a suitable unitary transformation $U_j$, $j=1,...,8$
according to Alice' results. In this way the initial state of
qubits 3 and 4 is restored on qubits 2 and 5.} \label{fig1}
\end{figure}

Let us consider the following situation.
Alice has at her side an unknown bipartite entangled state of the
form
\begin{equation}\label{bipartite}
|\varphi\rangle_{\sst 34}=\alpha |00\rangle_{\sst 34}+\beta|11\rangle_{\sst 34}
\:,
\end{equation}
and she wants to teleport this state to Bob. We assume that Alice
and Bob only share a single EPR pair (see Fig.\ref{fig1})
\begin{equation}\label{shared}
|\phi_{+}\rangle_{\sst 12}=\frac{1}{\sqrt{2}}\left\{|00\rangle_{\sst 12}+
|11\rangle_{\sst 12}\right\}\:
\end{equation}
Our protocol works as follows: Alice performs a Bell measurement
(BM) on qubits 1 and 3 and, after an Hadamard transformation,
measure qubit 4. Then she sends to Bob the results of both
measurements (overall three bits of classical information). Bob
introduces the additional qubit 5, and performs a C-not
transformation on qubits 2 and 5. Finally, he performs a unitary
transformation $U_j$, $j=1,...,8$ chosen accordingly to the
information received from Alice. In this way the initial state of
qubits 34 is restored on qubits 25.\par
\par\noindent
The initial state of the system is given by
\begin{equation}
|\psi\rangle_{\sst 12345}=|\phi_{+}\rangle_{\sst 12}|
\varphi\rangle_{\sst 34}|0\rangle_{\sst 5}\:,
\end{equation}
but, for the moment, let us consider only the first four qubits,
which are excited in the state $|\phi_{+}\rangle_{\sst 12}\:
|\varphi\rangle_{\sst 34}$. Alice performs a Bell measurement on
qubits $1$ and $3$, described by the projectors
$\Pi_{\phi_\pm}=|\phi_\pm\rangle_{\sst 13}{}_{\sst
13}\langle\phi_\pm |$ and $\Pi_{\psi_\pm}=|\psi_\pm\rangle_{\sst
13}{}_{\sst 13}\langle\psi_\pm |$, 
with
\begin{eqnarray}
|\phi_\pm \rangle_{\sst 13}=\frac{1}{\sqrt{2}} 
(|00\rangle_{\sst 13}\pm |11\rangle_{\sst 13}) \qquad 
|\psi_\pm \rangle_{\sst 13}=\frac{1}{\sqrt{2}} 
(|01\rangle_{\sst 13}\pm |10\rangle_{\sst 13})\nonumber 
\label{bellm}\;
\end{eqnarray}
According to her results, there are  four possible conditional
states of qubits $2$ and $4$, which are given by
\begin{eqnarray}
|\chi_{i}\rangle_{\sst 24}&=&\alpha
|00\rangle_{\sst 24}\pm\beta|11\rangle_{\sst 24}\qquad i=1,2 \label{chi1}\\
|\chi_{i}\rangle_{\sst 24}&=&
\beta|01\rangle_{\sst 24}\pm\alpha|10\rangle_{\sst 24}\qquad i=3,4
\label{chi2}\:.
\end{eqnarray}
The correspondence between the index $i$ and the result of
the BM is summarized in Table \ref{tab1}.
\begin{table}[h]
\caption{Summary of the eight possible combinations of Alice's
Bell (BM) and qubit 4 (Z$_4$) results, with the corresponding
conditional states of qubits $2$ and $4$, and conditional
transformations (CT) performed by Bob to restore the initial state
at his location. \label{tab1}}
\begin{center} $ $ \\ $ $ \\
\begin{tabular}{|c|c|c|c|c|c|}
\hline
& &BM$_{\sst 13}$$\quad$& CS $\varrho_{\sst 24}$$\quad$ & Z$_4$$
\quad$& Bob's CT$\quad$\\
     \cline{3-6}
$\;j=1$& $i=1$& $\phi_{+}$& $|\chi_{1}\rangle_{\sst 24}$& $0$&
${\mathbbm I}\otimes {\mathbbm I}$ \\
$\;j=2$& $i=1$& $\phi_{+}$& $|\chi_{1}\rangle_{\sst 24}$& $1$&
$\sigma_{z}\otimes {\mathbbm I}$ \\
$\;j=3$& $i=2$& $\phi_{-}$& $|\chi_{2}\rangle_{\sst 24}$& $0$&
$\sigma_{z}\otimes {\mathbbm I}$ \\
$\;j=4$& $i=2$& $\phi_{-}$& $|\chi_{2}\rangle_{\sst 24}$& $1$&
${\mathbbm I}\otimes {\mathbbm I}$ \\
$\;j=5$& $i=3$& $\psi_{+}$& $|\chi_{3}\rangle_{\sst 24}$& $0$&
$\sigma_{x}\otimes \sigma_{x}$ \\
$\;j=6$& $i=3$& $\psi_{+}$& $|\chi_{3}\rangle_{\sst 24}$& $1$&
$\sigma_{x}\otimes i\sigma_{y} $ \\
$\;j=7$& $i=4$& $\psi_{-}$& $|\chi_{4}\rangle_{\sst 24}$& $0$&
$\sigma_{x}\otimes i\sigma_{y}$ \\
$\;j=8$& $i=4$& $\psi_{-}$& $|\chi_{4}\rangle_{\sst 24}$& $1$&
$\sigma_{x}\otimes \sigma_{x}$ \\
\hline
\end{tabular} \end{center} \end{table}
\par\noindent
Let us now take into account also qubit $5$ at Bob's location, and
assume it is prepared in the state $|0\rangle_{\sst 5}$. At this
point, the global conditional state of the three qubits is given
by $|\chi_{i}\rangle_{\sst 24} |0\rangle_{\sst 5}$, $i=1,...,4$.
We assume that Bob performs a C-not transformation, denoted by the
unitary $C_{\sst 25}$, on subsystems $2$ and $5$, whereas Alice
rotates her qubit $4$ by a Hadamard transformation. The output
states resulting from this procedure are given by
$|\phi_i\rangle_{\sst 245}=C_{\sst 25}\otimes H_{\sst 4}\:
|\chi_{i}\rangle_{\sst 24}|0\rangle_{\sst 5}$. Explicitly we have
\begin{eqnarray}
|\phi_{\sst 1}\rangle_{\sst 245}&=&|0\rangle_{\sst 4}
\frac{1}{\sqrt{2}}\left\{\alpha
|00\rangle_{\sst 25}+\beta|11\rangle_{\sst 25}\right\}
+|1\rangle_{\sst 4}\frac{1}{\sqrt{2}}\left\{\alpha
|00\rangle_{\sst 25}-\beta|11\rangle_{\sst 25}\right\}
\nonumber\\
|\phi_{\sst 2}\rangle_{\sst 245}&=&|0\rangle_{\sst 4}
\frac{1}{\sqrt{2}}\left\{\alpha
|00\rangle_{\sst 25}-\beta|11\rangle_{\sst 25}\right\}
+|1\rangle_{\sst 4}\frac{1}{\sqrt{2}}\left\{\alpha
|00\rangle_{\sst 25}+\beta|11\rangle_{\sst 25}\right\}
\nonumber\\
|\phi_{\sst 3}\rangle_{\sst 245}&=&|0\rangle_{\sst 4}
\frac{1}{\sqrt{2}}\left\{\beta
|00\rangle_{\sst 25}+\alpha|11\rangle_{\sst 25}\right\}
-|1\rangle_{\sst 4}\frac{1}{\sqrt{2}}\left\{\beta
|00\rangle_{\sst 25}-\alpha|11\rangle_{\sst 25}\right\}
\nonumber\\
|\phi_{\sst 4}\rangle_{\sst 245}&=&|0\rangle_{\sst 4}
\frac{1}{\sqrt{2}}\left\{\beta
|00\rangle_{\sst 25}-\alpha|11\rangle_{\sst 25}\right\}
-|1\rangle_{\sst 4}\frac{1}{\sqrt{2}}\left\{\beta
|00\rangle_{\sst 25}+\alpha|11\rangle_{\sst 25}\right\}
\:,
\end{eqnarray}
where the subscript in $|\phi_i\rangle$ has the same meaning that
in Eqs. (\ref{chi1}) and (\ref{chi2}). Now Alice measures the
qubit $4$ and sends the results to Bob, together with the result
of the BM. The global amount of classical information from Alice
to Bob is thus three bits. Indeed, according to the results of the
BM and of the measurement of the qubit $4$, there are $8$ possible
conditional states at Bob's site. In order to restore the original
state, Bob should perform an appropriate unitary transformation
$U_j$ with $j=1,..,8$. It is straightforward to demonstrate that
the $U_j$ are the factorized transformations given in Table
\ref{tab1}. We have thus demonstrated that teleportation of
bipartite states of the form (\ref{bipartite}) is possible using a
single EPR pair shared
between Alice and Bob as far as an additional qubit and a C-not
transformation are available at the receiver's location. Three
bits of classical information should be sent from Alice to Bob in
order to restore the input state.\par
Notice that the choice of expression (\ref{bipartite}) for the
state to be teleported does not imply loss of generality. In fact,
any linear combination of the form
\begin{equation}
|\phi\rangle_{34}=\alpha |m_1,n_1\rangle + \beta |m_2,n_2 \rangle
\end{equation}
with $m_j,n_j=0,1$ can be brought to (\ref{bipartite}) by a unitary 
transformation at the sender' side and then it can be teleported 
by the same  protocol. Also
the choice of the state for the qubit $5$ at Bob's location is
arbitrary. In fact, any initial preparation $|\theta\rangle_{\sst
5}=a |0\rangle_{\sst 5} + b |1\rangle_{\sst 5}$ of qubit $5$ can
be seen as the action of a given unitary on the state
$|0\rangle_{\sst 5}$, {\em i.e.} $|\theta\rangle_{\sst 5} =
V_{\sst 5}|0\rangle_{\sst 5}$. In this case teleportation can be
restored if Bob performs the transformation $C^\prime_{\sst 25} =
C_{\sst 25}\, ({\mathbbm I}_{\sst 2} \otimes V_{\sst 5}^\dag)$
instead of the C-not. This degree of freedom will be exploited in
the next Section, where a whole class of teleportation protocols
will be introduced and characterized. Finally, we mention that
teleportation works equally well using a different Bell state,
either $|\phi_+\rangle_{\sst 12}$ or $|\psi_{\pm}\rangle_{\sst
12}$, as quantum channel.
\section{Beyond the C-not}
\label{s:uf} 
\begin{figure}[h]
\begin{center}
\includegraphics[width=0.7\textwidth]{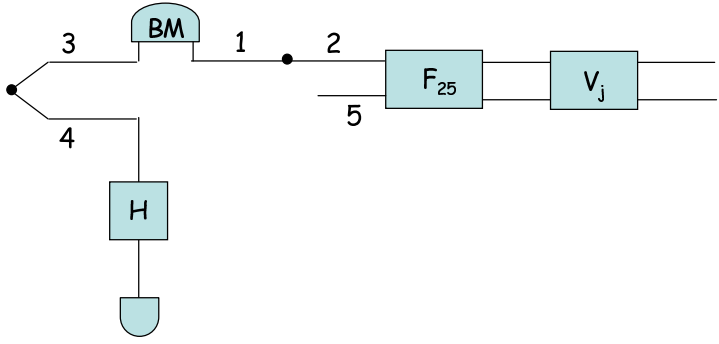}
\end{center}
\caption{Schematic diagram of a class of teleportation protocols.
Alice and Bob share a single EPR pair (qubit 1 and 2). Alice
performs a Bell measurement (BM) on qubit 1 and 3 and, after an
Hadamard transformation (H), measure qubit 4. Then she sends to
Bob the results of both measurements. Bob performs the two-qubit
transformation $F_{\sst 25}$ and then the unitary transformation
$V_j$, $j=1,...,8$ according to Alice' results. The initial state
of qubits 34 is restored on qubits 25 as far as $F_{\sst 25}$ has
nonzero entangling power.} \label{fig2}
\end{figure}

Let us consider a teleportation scheme similar to
that of Section \ref{s:cnot} where, instead of a C-not, Bob is
only able to perform a given generic two-qubit transformation
$F_{\sst 25}$ on qubits $2$ and $5$ (see Fig. \ref{fig2}). The
purpose of this Section is twofold. On one hand, we show that
teleportation can be still achieved by \textit{factorized}
conditional transformations at Bob site as far as $F_{\sst 25}$ ia
a nonzero entangling transformation. On the other hand, we show
that indeed there exists a class of transformations assuring the
success of the protocol, corresponding to a SU$(2)$ symmetry of
the set of transformations itself.
\par
We denote the action of $F_{\sst 25}$ on the standard basis by
$F_{\sst 25} |n,m\rangle = |f_{nm}\rangle$. Following the protocol
of Section \ref{s:cnot}, with $F_{\sst 25}$ instead of the C-not,
we have that after the BM and the measurement of qubit $4$, the
eight possible conditional states of qubits $2$ and $5$ are given
by
\begin{eqnarray}
|\theta_j\rangle_{\sst 25}&=& \alpha|f_{\sst 00}\rangle_{\sst 25} +
\beta|f_{\sst 11}\rangle_{\sst 25} \qquad j=1,4
\nonumber \\
|\theta_j\rangle_{\sst 25}&=& \alpha|f_{\sst 00}\rangle_{\sst 25} -
\beta|f_{\sst 11}\rangle_{\sst 25} \qquad j=2,3
\nonumber \\
|\theta_j\rangle_{\sst 25}&=& \beta |f_{\sst 00}\rangle_{\sst 25} +
\alpha|f_{\sst 11}\rangle_{\sst 25} \qquad j=5,8
\nonumber \\
|\theta_j\rangle_{\sst 25}&=& \beta |f_{\sst 00}\rangle_{\sst 25} -
\alpha|f_{\sst 11}\rangle_{\sst 25} \qquad j=6,7 \:.
\end{eqnarray}
\par\noindent
\par\noindent
Teleportation is possible if we find the appropriate unitaries $V_j$
to restore the initial state. The seeked transformations should fulfill
the following requirements
\begin{eqnarray}\label{defA}
j=1,4 \quad V_jF_{\sst 25}|00\rangle &=&|00\rangle   \quad
V_jF_{\sst 25}|10\rangle =|11\rangle  \\
j=2,3 \quad V_jF_{\sst 25}|00\rangle &=&|00\rangle   \quad
V_jF_{\sst 25}|10\rangle =-|11\rangle \label{defB} \\
j=5,8 \quad V_jF_{\sst 25}|00\rangle &=&|11\rangle   \quad
V_jF_{\sst 25}|10\rangle =|00\rangle  \\
j=6,7 \quad V_jF_{\sst 25}|00\rangle &=&|11\rangle  \quad
V_jF_{\sst 25}|10\rangle =-|00\rangle  \label{defF}\:.
\end{eqnarray}
Notice that these relations only determine two lines (the first and the fourth)
of the $4\times 4$ matrix representation of each $V_j$. Eqs. (\ref{defA}-\ref{defF})
thus imply
\begin{eqnarray}
V_4F_{\sst 25}=^{*}V_1F_{\sst 25}  \qquad
V_3F_{\sst 25}=^{*}V_2F_{\sst 25}  \label{eqV1}\\
V_8F_{\sst 25}=^{*}V_5F_{\sst 25}  \qquad V_7F_{\sst 25}=^{*}V_6F_{\sst 25} \label{eqV2}
\end{eqnarray}
where $=^*$ means equality {\em only} on the relevant lines.
Explicitly, Eqs.(\ref{defA}-\ref{defF}) can be rewritten as
\begin{eqnarray}
V_jF_{\sst 25} &=& \left(\begin{array}{cccc}
1 & 0 & 0 & 0 \\
z_{\sst j1} & z_{\sst j2} & z_{\sst j3} & z_{\sst j4} \\
z_{\sst j5} & z_{\sst j6} & z_{\sst j7} & z_{\sst j8} \\
0 & 0 & 1 & 0
\end{array} \right) \quad 
V_jF_{\sst 25} = \left(\begin{array}{cccc}
1 & 0 & 0 & 0 \\
z_{\sst j1} & z_{\sst j2} & z_{\sst j3} & z_{\sst j4} \\
z_{\sst j5} & z_{\sst j6} & z_{\sst j7} & z_{\sst j8} \\
0 & 0 & -1 & 0
\end{array} \right) \nonumber \\
V_jF_{\sst 25}&=& \left(\begin{array}{cccc}
0 & 0 & 1 & 0 \\
z_{\sst j1} & z_{\sst j2} & z_{\sst j3} & z_{\sst j4} \\
z_{\sst j5} & z_{\sst j6} & z_{\sst j7} & z_{\sst j8} \\
1 & 0 & 0 & 0
\end{array} \right) \quad 
V_jF_{\sst 25} = \left(\begin{array}{cccc}
0 & 0 & -1 & 0 \\
z_{\sst j1} & z_{\sst j2} & z_{\sst j3} & z_{\sst j4} \\
z_{\sst j5} & z_{\sst j6} & z_{\sst j7} & z_{\sst j8} \\
1 & 0 & 0 & 0
\end{array} \right)\;, \label{4mat}
\end{eqnarray}
where $z_{jk}$ are complex parameters and where, from top left to 
bottom right $j=1,4$, $j=2,3$, 
$j=5,8$ and $j=6,7$ respectively, Now, imposing the condition
of (special) unitarity to matrices in Eq. (\ref{4mat}), we obtain
the following expression for the second and third line of each
product $V_jF_{\sst 25}$
\begin{equation}
V_jF_{\sst 25}= \left(\begin{array}{cccc}
.. & .. & .. & .. \\
0  & \cos\gamma_j & 0 & -\sin\gamma_j\;e^{i(\theta_j -\phi_j )}  \\
0  & \sin\gamma_j\;e^{i\theta_j} & 0 & \cos\gamma_j\;e^{i\phi_j} \\
.. & .. & .. & ..\label{tre}
\end{array} \right)\:.
\end{equation}
The initial $8$ complex parameter $z_{jk}$ are thus reduced
to the three real phases $\gamma_j$, $\theta_j$ and $\phi_j$,
corresponding to a SU$(2)$ symmetry of the set $V_j$.
Once the transformation $F_{\sst 25}$ is known, Eqs. (\ref{4mat})
and (\ref{tre}) fully determine the form the transformations
$V_j$, each of them being determined in a class
with three real parameters. This degree of freedom
can be used to reduce the number of conditional transformations
needed to reconstruct the state nearby Bob, and to realize them
as factorized operations on the two qubits. For example, we can
choose $V_1=V_4$, $V_2=V_3$, $V_5=V_8$ and $V_6=V_7$, as it was
implicitly assumed in Section \ref{s:cnot}.
\par
Notice that factorized $V_j$, as it was in the scheme of Section
\ref{s:cnot}, can be obtained iff $F_{\sst 25}$ have a nonzero
entangling power {\em i.e.} full rank, otherwise Eqs. (\ref{4mat})
cannot be inverted, and no set of unitaries $V_j$ exists to
restore the initial state at Bob site. On the other hand, if we
accept that the $V_j$'s may be genuine two-qubit transformations,
then (\ref{4mat}) have solutions for a generic $F_{\sst 25}$. In
this case, however, the role of $F_{\sst 25}$ is irrelevant, and
it can be included in a redefinition of the $V_j$'s.
\section{Effect of noise}\label{s:noise}
Using the protocols described in the previous Sections, with a single
pure maximally entangled state as quantum channel, perfect teleportation
of bipartite states (\ref{bipartite}) may be achieved. However, entanglement can be in
general corrupted by the interaction with the environment. Therefore,
entangled states that are available for experiments are usually mixed
states, and it becomes crucial to establish whether or not the nonlocal
character of the protocols has survived the environmental noise.
\par
Teleportation is a linear operation, which also work with mixed
states. For a single qubit, quantum teleportation with a mixed
quantum channel was studied and it was shown that also when the
channel is not maximally entangled the fidelity is better than any
classical communication procedure \cite{popescu94}.
\par
\begin{figure}[h!]
\begin{center}
\includegraphics[width=0.35\textwidth]{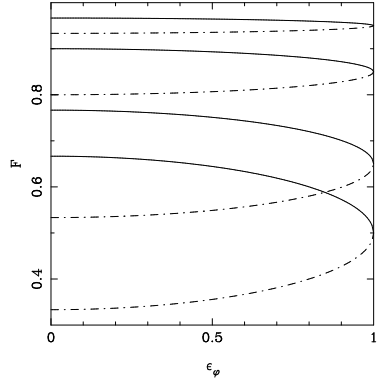}
\end{center}
\caption{Fidelity of teleportation as a function of the initial
entanglement $\epsilon_{\varphi}$ for different values of the
channel entanglement $\epsilon_c$. The solid and the dot-dashed
lines refer to the two branches $F_{\pm}$ respectively. From top
to bottom the curves are plotted for $\epsilon_c=0,9; 0,7; 0,3;
0.$ } \label{fig3}
\end{figure}

In this section we study the effect of noise on our schemes, and
analyze to which extent the teleportation of bipartite states is
degraded. Both fidelity of teleportation and entanglement of the
replica will be evaluated. As noisy quantum channel (qubits 1
and 2) we consider a Werner state of the form
\begin{equation}
\rho_c=p|\phi_+\rangle\langle \phi_+|+(1-p)\frac{\mathbb{I}\otimes
\mathbb{I}}{4} \label{werner}\:.
\end{equation}
The same results are obtained using a Werner state built from
a different Bell state, either $|\phi_-\rangle$ or $|\psi_\pm\rangle$.
As measure of entanglement for the bipartite state $\varrho$
we consider its "negativity"
\begin{equation}
\epsilon=-2\sum_{\lambda_{i}<0}\lambda_{i}\:,
\end{equation}
{\em i.e.} the (double) sum of the negative eigenvalues of the
partial transpose $\varrho^{\tau}$. With this definition the
quantum channel (\ref{werner}) shows a nonzero degree of
entanglement iff $p\geq 1/3$. We have
\begin{equation}
\epsilon_c = \frac{3p-1}{2};\label{epsc}
\end{equation}
from which we also obtain $p=(2\epsilon_c+1)/3$. For the state
$|\varphi\rangle$ of Eq. (\ref{bipartite}), {\em i.e.} the state
to be teleported, the negativity is given by
\begin{equation}
\epsilon_{\varphi} = 2|\alpha|\sqrt{1-|\alpha|^2}\:.\label{epsf}
\end{equation}
Notice that in this case we have $|\alpha|^2=\frac{1}{2}(1\pm
\sqrt{1-\epsilon_{\varphi}^2})$ i.e. the same degree of
entanglement is obtained for two different values of $|\alpha|^2$
symmetric with respect to $|\alpha|^2=1/2$. Using the mixed state
(\ref{werner}) as quantum channel our teleportation schemes lead
to the following output teleported state for qubits 2 and 5
\begin{equation}
\varrho_t=p|\varphi\rangle\langle \varphi|+(1-p)\frac{\mathbb{I}}{2}\otimes
|0\rangle \langle 0|. \label{teleW}
\end{equation}
At this point we calculate the fidelity of teleportation $F=\langle\varphi |
\varrho_t |\varphi\rangle$, which is given by
\begin{equation}
F=p+\frac{(1-p)|\alpha|^2}{2}\:,
\end{equation}
and in terms of the entanglement parameters
\begin{eqnarray}
F_{\pm}=\frac{3\epsilon_c+1}{4} \pm
\frac{\epsilon_c-1}{12}\sqrt{1-\epsilon_{\varphi}^2}.
\end{eqnarray}
In Fig. \ref{fig3} we show the fidelity as a function of the
initial entanglement $\epsilon_{\varphi}$ for different values of
the channel entanglement $\epsilon_c$. Notice that for each value
of $\epsilon_c$ fidelity shows two branches, corresponding to the
two different values of $|\alpha|^2$ that give the same
$\epsilon_{\varphi}$. By increasing the entanglement of the
channel the two branches approach each other, and for
$\epsilon_c=1$ we have $F=1$ for any value of
$\epsilon_{\varphi}$. The average fidelity over all possible input
states, {\em i.e.} over all possible values of $\alpha$ is given
by
\begin{eqnarray}
\overline{F}=\frac{(1+2p)}{3}\:.
\end{eqnarray}
\par
\begin{figure}[h]
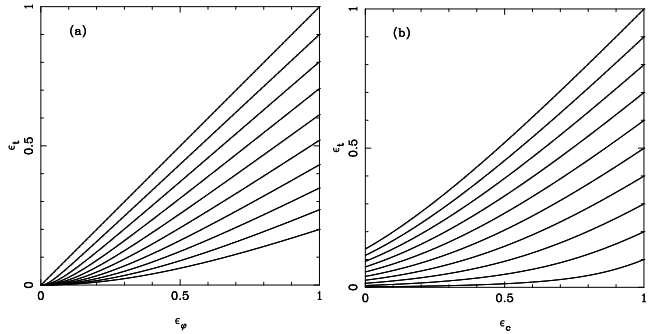

\begin{center}
\includegraphics[width=0.3\textwidth]{EtEf.ps}
\includegraphics[width=0.3\textwidth]{EtEc.ps}
\end{center}
\caption{(a): Entanglement of the teleported state $\epsilon_t$ as
a function of the initial entanglement $\epsilon_{\varphi}$ for
different values of the channel entanglement $\epsilon_c$ from $0$
to $1$ with a step $0.2$. (b): Entanglement of the teleported
state $\epsilon_\varphi$ as a function of the channel entanglement
$\epsilon_c$ for different values of the initial entanglement
$\epsilon_{\varphi}$ from $0$ to $1$ with a step $0.2$.}
\label{fig5} \label{fig4}
\end{figure}

\noindent
In order to evaluate the entanglement of the teleported state
$\varrho_t$ we calculated the eigenvalues of the partial
transpose. There is a single negative eigenvalue, given by
\begin{equation}
\lambda_{neg}=\frac{1}{4}\left[(1-p)-\sqrt{1+p[16|\alpha|^2(1
-|\alpha|^2)p+p-2]}\right] \:.
\end{equation}
Using the above expression together with Eqs. (\ref{epsc}) and
(\ref{epsf}) the entanglement of the replica state can be written
as
\begin{equation}
\epsilon_t=\frac{1}{3}\left\{\epsilon_c-1+\sqrt{(1-\epsilon_c)^2
+\epsilon_\varphi^2 (1+2\epsilon_c)^2}\right\}\:.
\end{equation}
In Fig. \ref{fig4} we show the entanglement $\epsilon_t$ of the
replica as a function of the initial entanglement
$\epsilon_{\varphi}$ for different values of channel entanglement
$\epsilon_c$, and as a function of the channel entanglement
$\epsilon_c$ for different values of the initial entanglement
$\epsilon_{\varphi}$. Notice that if the entanglement of the
initial state $\epsilon_{\varphi}$ is zero, then the entanglement
of the teleported state is always zero, independently on the
entanglement of the quantum channel. On the contrary, if
$\epsilon_{\varphi}$ is different from zero we have that
$\epsilon_t$ is different from zero also when $\epsilon_c$ is
zero. This is due to the main feature of our schemes, {\em i.e.}
the use of a single pair as quantum channel. In other words, for
any entangled $|\varphi\rangle$ the state (\ref{teleW}) remains
entangled also for the range of values $p<1/3$.
\section{Conclusions}\label{s:outro}
We have suggested a class of quantum protocols to teleport
bipartite entangled states of two qubits. Our schemes require a
single entangled pair shared by the two parties and the
transmission of three bits of classical information. Compared to
previous proposals, using two EPR pairs or tripartite entangled
states as quantum channels, our schemes require a reduced amount
of entanglement to achieve the same task. On the other hand, a
larger amount of classical information should be transmitted, and
the introduction of an additional qubit and of an entangling
transformation nearby Bob is required. Noisy quantum channels have
been considered and the effects on the teleportation fidelity and
on the entanglement of the replica have been evaluated.

\end{document}